\pdfoutput=1
\newcommand*{\ATLASLATEXPATH}{latex/}
\documentclass[cernpreprint,texlive=2016,txfonts,UKenglish]{latex/atlasdoc}
\usepackage[backend=biber]{\ATLASLATEXPATH atlaspackage}
\usepackage{\ATLASLATEXPATH atlasphysics}
\usepackage{\ATLASLATEXPATH atlasbiblatex}

\usepackage{\ATLASLATEXPATH atlascontribute}

\graphicspath{{logos/}{figures/}}

\usepackage{paper-defs}

\newcommand{\PbPb}{\mbox{Pb+Pb}}

\newcommand{\NUM}{\mbox{$N^\mathrm{ch}$}}
\newcommand{\MOM}{\mbox{$P_\mathrm{T}^\mathrm{ch}$}}

\newcommand{\Ncoll}{\mbox{$N_{\mathrm{coll}}$}}
\newcommand{\Ncone}{\mbox{$N_{\mathrm{cone}}$}}

\newcommand{\Njet}{\mbox{$N_{\mathrm{jet}}$}}
\newcommand{\Nch}{\mbox{$N_{\mathrm{ch}}$}}

\newcommand{\nchUE}{\mbox{$n_{\mathrm{ch}}^{\mathrm{UE}}$}}
\newcommand{\cnchUE}{\mbox{$\tilde{n}_{\mathrm{ch}}^{\mathrm{UE+fake}}$}}
\newcommand{\Npart}{\mbox{$N_{\mathrm{part}}$}}

\newcommand{\RFour}{\mbox{$R = 0.4$}}

\newcommand{\RTwo}{\mbox{$R= 0.2$}}

\newcommand{\TAA}{\mbox{$T_{\mathrm{AA}}$}}

\newcommand{\ETfcal}{\mbox{$\Sigma E_{\mathrm{T}}^{\mathrm{FCal}}$}}

\newcommand{\Dz}{\mbox{$D(z)$}}
\newcommand{\Dzmeas}{\mbox{$D^{\mathrm{meas}}(z, \pTjet)$}}
\newcommand{\Dzsub}{\mbox{$D^{\mathrm{sub}}(z, \pTjet)$}}
\newcommand{\Dzunf}{\mbox{$D^{\mathrm{unfolded}}(z)$}}

\newcommand{\Dptmeas}{\mbox{$D^{\mathrm{meas}}(\pTch, \pTjet)$}}
\newcommand{\Dptsub}{\mbox{$D^{\mathrm{sub}}(\pTch, \pTjet)$}}
\newcommand{\Dptunf}{\mbox{$D^{\mathrm{unfolded}}(\pT)$}}

\newcommand{\etajet}{\mbox{$\eta^{\mathrm{jet}}$}}
\newcommand{\phijet}{\mbox{$\phi^{\mathrm{jet}}$}}
\newcommand{\pTjet}{\mbox{$p_{{\mathrm{T}}}^{{\mathrm{jet}}}$}}

\newcommand{\pTch}{\mbox{$p_{{\mathrm{T}}}^{\mathrm{ch}}$}}
\newcommand{\pTtrk}{\mbox{$p_{{\mathrm{T}}}^{\mathrm{trk}}$}}
\newcommand{\pTrec}{\mbox{$p_{{\mathrm{T}}}^{\mathrm{rec}}$}}
\newcommand{\pTtrue}{\mbox{$p_{\mathrm{T}}^{\mathrm{truth}}$}}

\newcommand{\Rdz}{\mbox{$R_{D(z)}$}}
\newcommand{\Rdpt}{\mbox{$R_{D(\pT)}$}}

\newcommand{\diff}{\mathrm{d}}

\newcommand{\ETjet}{\mbox{$\ET^{\mathrm{jet}}$}}

\newcommand{\ptjet}{\mbox{$p_{\mathrm{T}}^{\mathrm{jet}}$}}

\newcommand{\ptparticle}{\mbox{$p_{\mathrm{T}}^{\mathrm{particle}}$}}

\renewcommand{\_}{{\tt \char`\_}}  % works properly only in \tt mode (!)

\newcommand{\ptmin}{\mbox{$p_{{\mathrm{T,min}}}$}}
\newcommand{\ptmax}{\mbox{$p_{{\mathrm{T,max}}}$}}
\newcommand{\fd}{\mathrm{d}}

\hypersetup{pdftitle={ATLAS document},pdfauthor={The ATLAS Collaboration}}

\begin{document}

\maketitle

%\tableofcontents

% List of contributors - print here or after the Bibliography.
%\PrintAtlasContribute{0.30}
%\clearpage

%%%%%%%%%%%%%%%%%%%%%%%%%%%
\section{Introduction}
\label{sec:intro}

Heavy-ion collisions at ultra-relativistic energies produce a medium 
of strongly interacting nuclear matter composed of deconfined colour
charges which is commonly called a quark--gluon plasma (QGP)
\cite{Adcox:2004mh,Adams:2005dq,Back:2004je,Arsene:2004fa}.
Hard-scattering processes occurring in these collisions produce
high transverse momentum, \pt , partons that propagate through the
medium and lose energy.
This phenomenon is termed ``jet quenching''. More specifically, 
jet quenching is a process in which constituents of the parton shower 
may be elastically or inelastically scattered by the constituents of the 
plasma, resulting in the suppression of jet production and the modification 
of the internal structure of jets
\cite{Mehtar-Tani:2013pia,Qin:2015srf,Blaizot:2015lma}. Inclusive-jet 
suppression has been measured previously at the LHC in terms of the 
nuclear modification factor \cite{Aad:2012vca,Abelev:2013kqa,Aad:2014bxa,Adam:2015ewa,Khachatryan:2016jfl}. 
A suppression of jet production by about a factor of two in central 
heavy-ion collisions was observed.
The internal structure of jets was also measured
\cite{Aad:2014wha,Chatrchyan:2013kwa,Chatrchyan:2014ava,Adam:2015doa} and these measurements revealed 
modification of the distributions of the jet fragments.
  The measurements of the jet structure were
supplemented by a measurement of the correlation of the jet 
suppression with missing transverse momentum
\cite{Chatrchyan:2011sx}, leading to a conclusion that the energy
lost by partons is transferred predominantly to soft particles being
radiated at large angles with respect to the direction of the
original parton.

This paper presents a new measurement of the internal structure of jets by 
ATLAS in \PbPb\ and \pp\ collisions, both at the same centre-of-mass 
energy per colliding nucleon pair of 2.76~\TeV.
The measurement utilised \PbPb\ data collected during 2011 
corresponding to an integrated luminosity of 0.14~nb$^{-1}$ as well as 
data from \pp\ collisions recorded during 2013 corresponding to 4.0~pb$^{-1}$.
In this paper the same quantities that were introduced in Ref.~\cite{Aad:2014wha} are used,
namely the jet fragmentation functions, $D(z)$, and distributions of 
charged-particle transverse momenta measured inside the jet, $D(\pt)$. The 
$D(z)$ distributions are defined as
  \begin{equation}
\Dz \equiv \dfrac{1}{\Njet} \dfrac{\diff\Nch}{\diff z},
\label{eq:ffuncdef}
  \end{equation}
where $\Njet$ is the total number of jets, \Nch\ is the number of charged particles associated with a jet, and the longitudinal momentum fraction $z$ is defined as
  \begin{equation}
z \equiv   \frac{\pt}{\ptjet} \cos \Delta R  
= \frac{\pt}{\ptjet} \cos {\sqrt{(\Delta \eta)^2 + (\Delta \phi)^2}}.
\label{formula:z}
  \end{equation}
Here \ptjet\ is the transverse momentum of a jet measured with respect to the beam direction, \pt\ stands for the transverse momentum of a charged particle, 
$\Delta \eta$ and $\Delta \phi$ are the distance between the jet axis 
and the charged-particle direction in pseudorapidity and 
azimuth,\footnote{ATLAS uses a  
right-handed coordinate system with its origin at the nominal
interaction point (IP) in the centre of the detector and the
$z$-axis along the beam pipe. The $x$-axis points from the IP to the
centre of the LHC ring, and the $y$-axis points upward. Cylindrical
coordinates $(r,\phi)$ are used in the transverse plane, $\phi$
being the azimuthal angle around the beam pipe. The pseudorapidity
is defined in terms of the polar angle $\theta$ as
$\eta=-\ln\tan(\theta/2)$. 
Rapidity is defined as $y = 0.5 \ln \frac{E+p_z}{E-p_z}$ where
$E$ and $p_z$ are the energy and the component of the momentum along the beam direction.} respectively.
  \footnote{The $\Delta R \equiv \sqrt{(\Delta \eta)^2 + (\Delta \phi)^2}$ used here is a boost-invariant 
replacement for the polar angle $\theta$ between the jet and charged particle.}
  The $D(\pt)$ distributions are defined as
  \begin{equation}
D(\pt) \equiv \frac{1}{\Njet} 
\frac{\fd \Nch(\pT)}{\fd \pT} .
  \end{equation}
The fragmentation distributions were measured for jets reconstructed 
with the \antikt\ algorithm \cite{Cacciari:2008gp} with the radius 
parameter set to $R=0.4$. The charged particles were matched to a jet 
by requiring the distance between the jet axis and the charged 
particle to be $\Delta R < 0.4$.
The fragmentation distributions were fully corrected to the particle level.

In the first measurement of jet fragmentation by ATLAS in heavy-ion collisions
~\cite{Aad:2014wha}, the measurements were performed for jets 
with the radius parameters $R=0.2,0.3,$ and 0.4. Jet fragments having a minimum 
\pt\ of $2~\GeV$ were measured within an angular range $\Delta R = 0.4$ 
from the jet axis. The $D(z)$ and $D(\pt)$ distributions were presented 
for seven bins in collision centrality. 
Ratios of fragmentation functions in the different centrality bins to the 
\mbox{60--80\%} bin were presented and used to evaluate 
the modifications of the jet fragmentation caused by the medium. Those ratios exhibited an 
enhancement in fragment yield in central collisions for $z \lesssim 
0.04$, a reduction in fragment yield for $0.04 \lesssim z \lesssim 0.2$, 
and an enhancement in the fragment yield for $z > 0.4$. The 
modifications were found to decrease monotonically with decreasing 
collision centrality from \mbox{0--10\%} to \mbox{50--60\%}. A similar 
set of modifications was observed in the $D(\pt)$ distributions 
over corresponding \pT\ ranges.

This new analysis provides a measurement of the jet structure of \RFour\ jets using 
the same observables, but it decreases the minimum \pt\ for charged particles to 1~\GeV\ 
and evaluates the fragmentation observables differentially in jet 
\pt\ and $y$. Furthermore, the new analysis uses the fragment 
distributions measured in \pp\ collisions as a reference for the 
measurement of jet fragmentation in heavy-ion collisions. Using this  
information about the jet structure, the 
flow of the quenched jet energy and number of charged particles was quantified as a function of the centrality.

The content of this paper is organised as follows: 
Section~\ref{sec:experiment} describes the experimental set-up.
Section~\ref{sec:datasets} describes the event selection and data sets.
The jet and track reconstruction and selection are introduced in Section~\ref{sec:reconstruction}.
Section~\ref{sec:analysis} discusses the analysis procedure.
The estimation of systematic uncertainties is given is Section~\ref{sec:systematics}.
Section~\ref{sec:results} describes the results of the measurement.
Section~\ref{sec:discussion} provides a discussion of the results, and 
Section~\ref{sec:summary} summarises the analysis.

%%%%%%%%%%%%%%%%%%%%%%%%%%%
\section{Experimental set-up}
\label{sec:experiment}

The measurements presented in this paper were performed using the ATLAS 
calorimeter, inner detector, trigger, and data acquisition systems 
\cite{Aad:2008zzm}. The ATLAS calorimeter system consists of a liquid argon (LAr) 
electromagnetic (EM) calorimeter covering $|\eta|<3.2$, a steel--scintillator 
sampling hadronic calorimeter covering $|\eta| < 1.7$, a LAr hadronic calorimeter 
covering $1.5 < |\eta| < 3.2$, and a LAr forward calorimeter (FCal) covering 
$3.2 < |\eta| < 4.9$. 
The hadronic calorimeter has three sampling layers, longitudinal in shower depth, and has a 
$\Delta \eta \times \Delta \phi$ granularity of $0.1 \times \pi/32$ for $|\eta| < 
2.5$ and $0.2 \times 2\pi/32$ for $2.5 < |\eta| < 4.9$.\footnote{
Except the third sampling layer, which has a segmentation of $0.2 \times \pi/32$
up to $|\eta| = 1.7$.}
The EM calorimeters are longitudinally segmented in shower depth into three compartments with an 
additional pre-sampler layer. The EM calorimeter has a granularity that varies 
with layer and pseudorapidity, but which is generally much finer than that of the 
hadronic calorimeter. The middle sampling layer, which typically has the largest 
energy deposit in EM showers, has a granularity of $0.025 \times 0.0245$ over 
$|\eta| < 2.5$.

The inner detector \cite{Collaboration:2010knc} measures charged particles within 
the pseudorapidity interval $|\eta|<2.5$ using a combination of silicon pixel 
detectors, silicon microstrip detectors (SCT), and a straw-tube transition radiation 
tracker (TRT), all immersed in a 2~T axial magnetic field. All three detectors are 
composed of a barrel and two symmetrically placed end-cap sections. The pixel 
detector is composed of three layers of sensors with nominal feature size $50~\mathrm{ 
\mu m} \, \times \, 400~\mathrm{\mu m}$.  The microstrip detector's barrel section contains 
four layers of modules with 80~$\mu \mathrm{m}$ pitch sensors on both sides, while 
the end-caps consist of nine layers of double-sided modules with radial strips 
having a mean pitch of $80~\mathrm{\mu m}$. The two sides of each layer in both 
the barrel and the end-caps have a relative stereo angle of 40~mrad. The 
transition radiation tracker contains up to 73 (160) layers of staggered straws 
interleaved with fibres in the barrel (end-cap). Charged particles with $\pT 
\gtrsim 0.5$~\GeV\ and $|\eta| < 2.5$ typically traverse three layers of 
silicon pixel detectors, four layers of double-sided microstrip sensors, and 36 
straws if $|\eta| < 2.0$.

Minimum-bias \PbPb\ collisions were selected using measurements 
from the zero-degree calorimeters (ZDCs) and the minimum-bias trigger 
scintillator (MBTS) counters \cite{Aad:2008zzm}. The ZDCs are located 
symmetrically at a longitudinal distance of~$\pm 140$~meters from the detector centre 
and cover $|\eta| > 8.3$. In \PbPb\ 
collisions, the ZDCs primarily measure ``spectator'' neutrons, which 
originate from the incident nuclei and do not interact hadronically.
The MBTS detects charged particles over $2.1 < |\eta| < 3.9$ using 
two counters placed at 
a distance of $\pm 3.6$~m from the interaction point. 
Each counter is divided into 16 modules with 8 different positions 
in $\eta$ and~$\phi$. 
Each counter provides measurement of both the pulse heights and 
arrival times of ionisation energy deposits.

%%%%%%%%%%%%%%%%%%%%%%%%%%%
\section{Event selection and data sets}
\label{sec:datasets}

The analysis utilised  an integrated luminosity of 0.14~nb$^{-1}$ of Pb+Pb data and 
4.0~pb$^{-1}$ of \pp\ data collected in 2011 and 2013, respectively. 
The Pb+Pb events used in the analysis 
were required to have a reconstructed primary vertex and a time difference 
between hits from the two sides of the MBTS detector of less than 3~ns. The primary 
vertices were reconstructed from charged-particle tracks with \mbox{$\pT > 
0.5~\GeV$}. The tracks were reconstructed from hits in the inner detector using the 
standard track-reconstruction algorithm \cite{Cornelissen:2008zzc} with settings 
optimised for the high hit density in heavy-ion collisions \cite{ATLAS:2011ah}. 
The Pb+Pb events were selected for recording by a combination of Level-1 
minimum-bias and High Level Trigger (HLT) jet triggers.
The Level-1 trigger required a total transverse energy measured in the 
calorimeter of greater than 10~\GeV. The HLT jet trigger ran the offline \PbPb\ 
jet-reconstruction algorithm, described below, for \RTwo\ jets except for the 
application of the final hadronic energy-scale correction. 
The HLT selected events containing an \RTwo\ jet with transverse energy $\ET > 20$~\GeV\ in the $|\eta| < 3.2$ range.
A total of 14.2 million events satisfied these event selection criteria. 
The performance of the jet triggering is summarised in 
Ref.~\cite{ATLAS:2016qun}. 

  The centrality of \PbPb\ collisions was characterised by \ETfcal, the 
total transverse energy measured in the forward calorimeters (FCal) 
\cite{ATLAS:2011ah}. The results in this paper were obtained using 
seven centrality bins defined according to successive percentiles of the 
\ETfcal\ distribution ordered from the most central, highest \ETfcal , to the most 
peripheral collisions: 0--10\%, 10--20\%, 20--30\%, 30--40\%, 40--50\%, 
50--60\%, and 60--80\%. The percentiles were defined after correcting the 
\ETfcal\ distribution for the 2\% minimum-bias trigger inefficiency 
 which only affects the most peripheral collisions (80--100\%), that were not included in this analysis.

The \pp\ events used in the analysis were selected using the ATLAS jet 
trigger~\cite{Aad:2012xs} with a requirement of a minimum jet~\pt\ of 75~\GeV.   
The \pp\ events were required to contain at least one primary vertex, 
reconstructed from at least two tracks with $\pt>0.5~\GeV$. Jets 
originating from all selected events were included in the measurement.

The performance of the ATLAS detector and offline analysis in 
measuring jets and charged particles in \pp\ collisions was evaluated 
using a sample of 15 million Monte Carlo (MC) events obtained from 
PYTHIA~\cite{Sjostrand:2006za} hard-scattering events (using PYTHIA version 
6.425, with parameter values set to the AUET2B tune~\cite{ATLAS-PHYS-PUB-2011-009},
and CTEQ6L1 parton distribution 
functions \cite{Pumplin:2002vw}).
The generator-level spectrum of $R=0.4$ jets covers the 
transverse momentum interval of $20 < \pt < 500~\GeV$, which is sufficient to cover the jet \pt\ range in the data.  The detector 
effects were fully simulated \cite{atlassim} using GEANT4 
\cite{Agostinelli:2002hh}.
The reconstruction performance in \PbPb\ collisions 
was evaluated using a sample of 18 million events obtained by overlaying 
simulated PYTHIA hard-scattering events onto minimum-bias \PbPb\ events 
recorded in 2011. In this overlay procedure, the simulated hits were 
combined with the data from minimum-bias events to produce the final 
sample. The generator-level spectrum of jets in the overlay sample covers 
the transverse momentum interval of $35 < \pt < 560~\GeV$.
In all samples, the generator-level charged particles are defined as 
all final-state charged PYTHIA particles with lifetimes longer than $0.3 \cdot 
10^{-10}$~s originating from the primary interaction or from the subsequent 
decay of particles with shorter lifetimes.\footnote{While generator-level 
charged particles are massive, the tracks reconstructed in the inner 
detector are massless.}

%%%%%%%%%%%%%%%%%%%%%%%%%%%
\section{Jet and track selection}
\label{sec:reconstruction}

  Jets were reconstructed using the techniques described in 
Ref.~\cite{Aad:2012vca}, which are briefly summarised here.
The \antikt\ $R=0.4$ algorithm was first run in four-momentum 
recombination mode on calorimeter cells grouped into $\Delta \eta \times \Delta \phi = 0.1\times 0.1$ 
calorimeter towers. The tower kinematics 
were obtained by summing electromagnetic-scale energies \cite{Aad:2011he} of massless
calorimeter cells within the tower boundaries. 
In the case of the reconstruction of jets in \PbPb\ collisions, an 
underlying event (UE) subtraction was performed in the following way.
An iterative procedure was used to estimate a layer-dependent 
and pseudorapidity-dependent UE energy density while 
excluding jets from that estimate. The UE energy was corrected 
for the presence of the elliptic flow \cite{Poskanzer:1998yz}, which was 
subtracted from each calorimeter cell within the towers included in the 
reconstructed jet. The final jet kinematics were calculated via a 
four-momentum sum of all cell energy deposits (assumed massless) contained within the jet. The UE 
contribution was subtracted at the cell level. A correction was applied 
to the reconstructed jet to account for jets not excluded or only 
partially excluded from the UE estimate. Finally, the jet $y$- and 
\ET-dependent hadronic energy-scale calibration factor was applied in both the 
\pp\ and \PbPb\ collisions.

In the trigger the HLT reconstruction algorithms described in Ref.~\cite{ATLAS:2016qun} were used.
The HLT jet trigger selection is fully efficient 
at a \pt\ of approximately 90~\GeV. This, together with the intention to provide the results in the 
jet \pt\ selections that are the same as bins used in 
Ref.~\cite{Aad:2014bxa}, limits the results to jets with $\pt > 
100~\GeV$. The jet reconstruction performance is
described in Ref.~\cite{Aad:2012vca}. In order to evaluate the 
rapidity dependence of the jet 
structure, jets were categorised in four rapidity 
intervals, namely $|y|<0.3, 0.3<|y|<0.8, 1.2<|y|<2.1$, and 
$|y|<2.1$. The rapidity interval of $0.8<|y|<1.2$ was not
considered in the analysis 
since the jet shape measurements are degraded in this region due to the 
transition in the detector between the SCT barrel and end-caps. 
  
The tracks from \pp\ collisions were required to have 
at least one hit in the pixel detector and six hits in the silicon 
microstrip detector. 
In order to reject secondary particles, the transverse ($d_0$) and 
longitudinal ($z_0\sin{\theta}$) impact parameters of the tracks 
measured with respect to the primary vertex were required to be smaller than %to be within 
1.5~mm (0.2~mm for $d_0$ if $\pt > 10~\GeV$).

In \PbPb\ collisions, the occupancies of the three tracking subsystems 
reached different levels. The pixel detector occupancy was below 1\% 
even in the most central collisions. The corresponding number for the 
SCT detector was below 10\%, while the occupancy in the TRT reached 90\% \cite{Aad:2015wga}. 
To account for the high occupancy in \PbPb\ events, the track 
reconstruction was configured differently from that in \pp\ collisions. 
Tracks from \PbPb\ collisions were required to have at least two hits 
in the pixel detector, including a hit in the first pixel layer if the 
hit was expected from the track trajectory,
and seven hits in the silicon microstrip detector.
In addition, the $d_0$ and $z_0\sin{\theta}$ of the tracks measured 
with respect to the primary vertex were required to satisfy 
$|d_0/\sigma_{d_0} | < 3$ and $|z_0\sin{\theta}/\sigma_{z}| <3$, where 
$\sigma_{d_0}$ and $\sigma_{z}$ are uncertainties on $d_0$ and 
$z_0\sin{\theta}$, respectively, obtained from the track-fit covariance 
matrix. All tracks used in this analysis were required to have $\pt > 
1~\GeV$.

The efficiency for reconstructing charged particles within jets was 
evaluated separately for \pp\ and \PbPb\ collisions using MC 
events, described in Section~\ref{sec:datasets}.  
The efficiency was evaluated for charged particles that satisfy the selection criteria 
described above and were matched to generator-level (``truth'') jets with $\pt > 100~\GeV$ 
in each of the four jet rapidity intervals. In the case of \PbPb\ 
collisions, the efficiency was evaluated separately for each 
centrality bin.
 
\begin{figure*} [ht]
\centerline{
\includegraphics[width=0.49\textwidth]{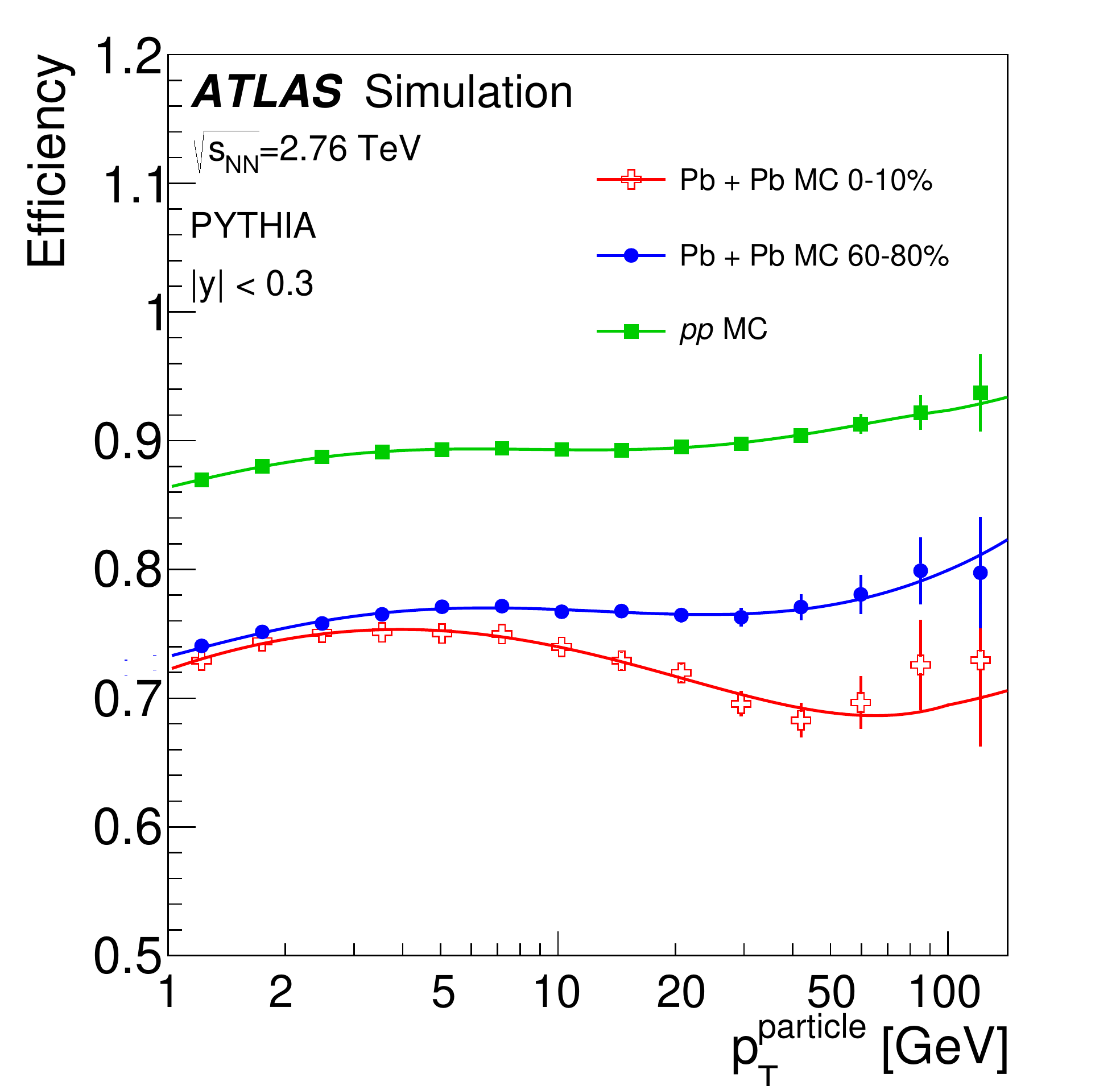} 
\includegraphics[width=0.49\textwidth]{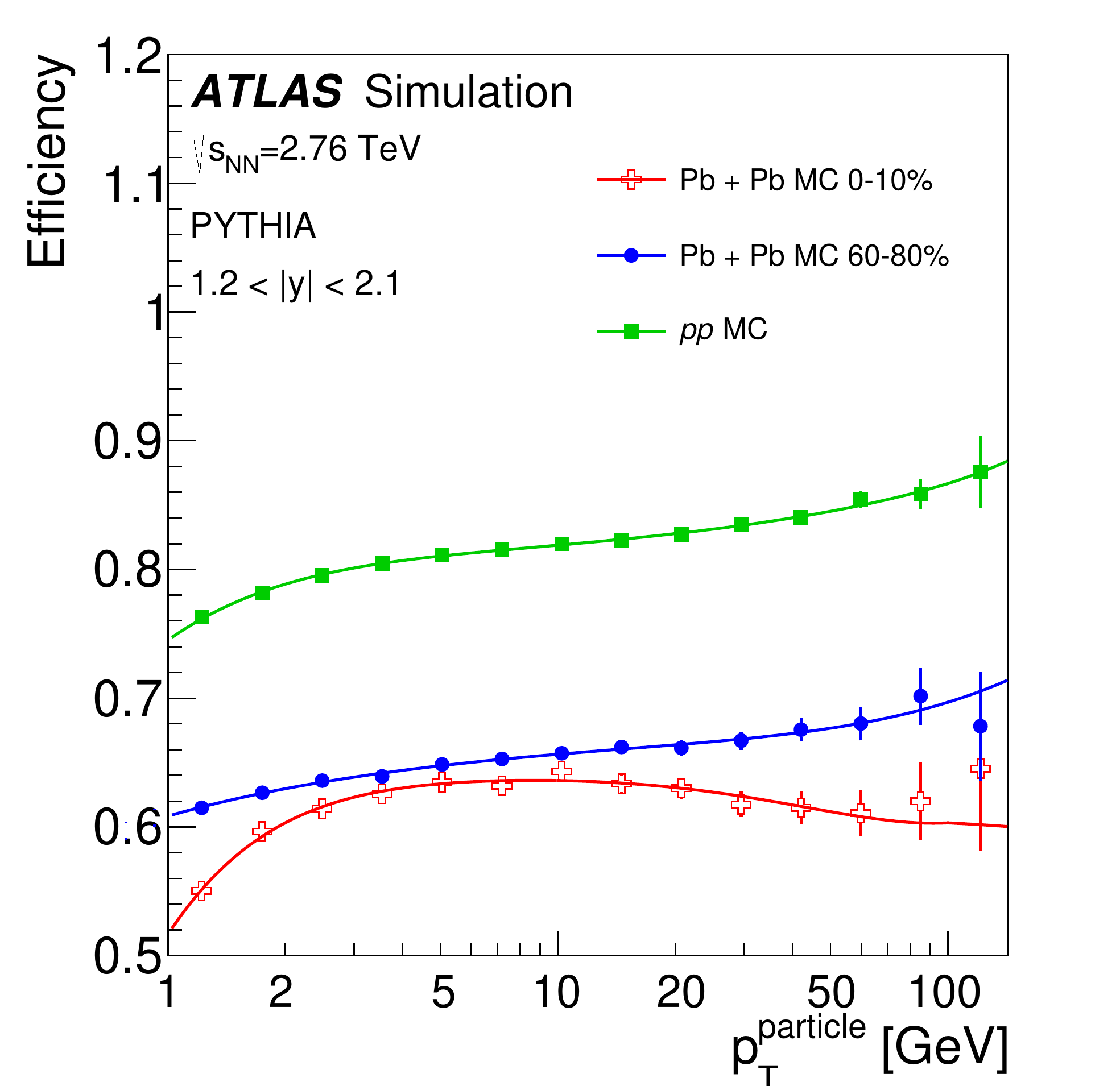}
}
\caption{
The tracking efficiency evaluated in simulation for particles in jets with $\ptjet > 
100~\GeV$ as a function of truth charged-particle transverse momentum, \ptparticle , for jets with $|y| < 0.3$ (left) and $1.2 < 
|y| < 2.1$ (right). Efficiency is shown for central and peripheral 
\PbPb\ collisions as well as for \pp\ collisions. The full line represents the 
parameterisation (for more details see the body of the text).
}
\label{fig:trackingeff}
\end{figure*}

The tracking efficiency correction $1/\varepsilon$ was evaluated as a function 
of charged-particle \pt\ and $y$.  
The tracking efficiency $\varepsilon$ was obtained as a ratio of tracks that have an 
associated truth charged particle to all the truth charged particles.
To guarantee smooth behaviour of the correction factors as a function 
of track \pt , the tracking efficiency was parameterised in the 
region of $1 < \pt < 90~\GeV$ using a fourth-order polynomial in
the logarithm of the track \pt. 
This functional form gives a good description of the onset of the efficiency at 
low~\pt\ as well as the behaviour in the intermediate-\pt\ region. At the 
same time it is not susceptible to statistical fluctuations in these 
regions. 
However, in the region of $\pt>90~\GeV$ the polynomial in the logarithm 
does not provide a good parameterisation of efficiencies.
The study of the high-\pt\ behaviour in both the 
\pp\ and \PbPb\ simulations showed that the tracking efficiency 
generally continues to follow the linear trend present at $\pt\ < 90~\GeV$. 
Thus, the result of the fit using a polynomial in the logarithm for tracks with 
$\pt>90~\GeV$ was replaced by a linear function with the slope 
determined from the difference between the fitted efficiencies at $\pt=70~\GeV$ and 
$\pt=90~\GeV$. 
The value of the slope does not exceed 0.001.
The efficiency for reconstructing tracks along with its 
parameterisation is shown in Figure~\ref{fig:trackingeff}.
The fake-track contribution was evaluated by matching 
reconstructed tracks to truth MC particles and found to be smaller than 2\%
for tracks satisfying the selection requirements defined above.

%%%%%%%%%%%%%%%%%%%%%%%%%%%
\section{Analysis procedure}
\label{sec:analysis}

  The analysis procedure is described briefly as follows. First, the 
measured distributions were corrected for the presence of a UE 
contribution (in the case of \PbPb\ collisions only) and for fake 
tracks. The corrected distributions were then unfolded using a 
two-dimensional Bayesian unfolding to correct for finite jet energy 
resolution and smearing due to finite track momentum resolution. The 
unfolded distributions were then normalised by the respective number of 
jets, which was obtained using one-dimensional Bayesian unfolding of jet 
\pt\ spectra. Details of each step in this procedure are discussed in 
the next paragraphs.

The first step in the analysis was to obtain measured two-dimensional uncorrected 
fragmentation functions, \Dzmeas , and the two-dimensional distribution of charged-particle 
transverse momenta measured inside the jet, \Dptmeas , which are 
defined using the following formulae,
\begin{eqnarray}
\Dptmeas \equiv \frac{1}{\varepsilon(\pTch, y)} \frac{\Delta \Nch(\pTch, \pTjet)}{\Delta \pTch}, \label{eq:Dptmeas} \\
\Dzmeas  \equiv \frac{1}{\varepsilon(\pTch, y)} \frac{\Delta \Nch(z, \pTjet)}{\Delta z}. \label{eq:Dzmeas} 
\end{eqnarray}
Here $\Delta \Nch(\pTch)$ and $\Delta \Nch(z)$ represent the number 
of measured charged particles within $\Delta R = 0.4$ of the jet axis
obtained from the \antikt\ clustering
in given bins of charged-particle transverse momentum, $\pTch$, and $z$ 
respectively.\footnote{The labels `ch' and `jet' are
used here to better distinguish the quantities 
connected with charged particles from quantities connected with jets.}
The variable $\varepsilon$ is the MC-evaluated track 
reconstruction efficiency.
The superscript `meas' in Eqs.~(\ref{eq:Dptmeas}), (\ref{eq:Dzmeas}) 
indicates that the measured distributions were corrected only 
for the tracking efficiency.
The corrections for the UE and detector effects were applied in the 
subsequent steps of the analysis as discussed in the next paragraphs.
   
Charged particles from the UE constitute a background that
needs to be subtracted from the measured distributions. This background 
depends on \pTch\ and $\eta^{\mathrm{ch}}$ of the charged particle, and the centrality of the 
collision.
The contribution of the UE background was evaluated for each measured 
jet using a grid of $\Delta R = 0.4$ cones that spanned the full coverage of 
the inner detector. The cones had a fixed distance between their 
centres chosen such that the coverage of the inner detector was
maximised while the cones did not overlap each other. 
To avoid biasing the UE estimate, cones associated with real jets have 
to be removed.
This was done by removing cones having a charged particle with 
$\pTch>6$~\GeV\ or having a distance $\Delta R < 0.4$ between its centre and the nearest 
jet with $\pt>90$~\GeV.

The resulting UE charged-particle yields, $\fd \nchUE/ \fd \pTch$ or 
$\fd \nchUE/ \fd z$, 
were evaluated over $1 < \pTch < 6$~\GeV\ as a function of
charged-particle \pTch, \pTjet , and \etajet\ and
averaged over all cones according to:
\begin{eqnarray}
\label{eq:background1}
\dfrac{\fd \nchUE}{ \fd \pTch} &=&
\dfrac{1}{\Ncone}
\dfrac{1}{\varepsilon}
\dfrac{\Delta N_{\mathrm{ch}}^{\mathrm{cone}}(\pTch,\pTjet,
\etajet)}{\Delta \pTch}, \\
\label{eq:background2}
\dfrac{\fd \nchUE}{\fd z} &=&
\dfrac{1}{\Ncone}
\dfrac{1}{\varepsilon}
\dfrac{\Delta N_{\mathrm{ch}}^{\mathrm{cone}}(z,\pTjet,
\etajet)}{\Delta z} \Bigg|_{z = \frac{
p_\mathrm{T}^\mathrm{ch}}{ p_\mathrm{T}^\mathrm{jet}} 
\cos \Delta R }.
\end{eqnarray}
Here \Ncone\ represents the number of background cones associated with 
a given jet with  \pTjet\ and \etajet , 
$\Delta N_{\mathrm{ch}}^{\mathrm{cone}}$ 
is the number of charged particles summed over all cones 
associated with the jet in question,
and $\Delta R$ represents the distance between the centre of a cone 
and the direction of a given charged particle. Not shown in 
Eqs.~(\ref{eq:background1}) and (\ref{eq:background2}) are correction 
factors that were applied to each background cone to correct for the 
difference in the average UE particle yield at a given \pTch\ between 
the $\eta$ position of the cone and \etajet\ and separate correction 
factors to account for the difference in the elliptic flow modulation at 
the $\phi$ position of the UE cone and \phijet. 
The former correction was based on a 
parameterisation of the \pTch\ and centrality dependence of charged-particle 
yields in minimum-bias collisions.
The latter correction was 
based on a parameterisation of the \pTch\ and centrality dependence of 
elliptic flow coefficients, $v_2$, measured by ATLAS 
\cite{ATLAS:2011ah}. 
Since the measurement was not performed with respect to the reaction 
plane, the impact of the flow correction was at the level of a few percent 
 of the UE yields.
By evaluating the UE yields only from events containing jets
included in the analysis, the background automatically had the correct
distribution of centralities within a given centrality bin.  

The UE yields need to be further corrected for the correlation 
between the actual UE yield in the jet and a finite, 
centrality-dependent jet energy resolution.
Due to the steeply falling \pt\ distribution of jets, the 
smearing due to jet energy resolution leads to a net migration of jets 
from lower \pt\ to higher \pt\ values (hereafter referred to as 
``upfeeding'') such that a jet reconstructed with a given \pTrec\  
corresponds, on average, to a truth jet with lower transverse momentum, \pTtrue.
The upfeeding was observed in the MC simulation to induce a difference between 
the determined UE yields, as described above, 
and the actual UE 
contribution to reconstructed jets. This difference was found to be  
centrality dependent, and it also exhibited a weak \pTjet\ dependence.
That difference was found to result from
intrinsic correlations between 
the UE contribution to the yield of particles measured inside the jet
and the MC \pTjet\ shift, $\Delta p_{\mathrm{T}}^{\mathrm{jet}}
= \pTrec - \pTtrue$.  In particular, jets with positive (negative)
$\Delta p_{\mathrm{T}}^{\mathrm{jet}}$ were found to have an UE
contribution larger (smaller) than jets with $\Delta
p_{\mathrm{T}}^{\mathrm{jet}} \sim 0$.  Due to the net upfeeding in
the falling jet spectrum, the selection of jets above a given
\pTjet\ threshold causes the UE contribution to be larger than that
estimated from the procedure described above. The average
fractional mismatch in the estimated UE background was found to
have a minor dependence on \pTch\ and \pTjet\ and to vary with
centrality by factors of 0--20\% with respect to the original
UE estimates.
To correct for this effect,
multiplicative correction factors, dependent on centrality, 
$y^\mathrm{jet}$, $\pTjet$ and $\pTch$ (or $z$) were applied
to the $\fd \nchUE/ \fd \pTch$ (or $ \fd \nchUE/ \fd z$) distributions. 
These multiplicative factors were estimated in MC samples as a ratio of UE 
distributions calculated from tracks within the area of a jet which do 
not have an associated truth particle and the UE distributions estimated by the cone method.
The measured distributions were also corrected for the presence of fake 
tracks by subtracting the fake-track contribution estimated in MC 
simulations.
The corrected UE distributions,
$\fd \cnchUE / \fd \pTch$ and 
$\fd \cnchUE / \fd z$, 
were then subtracted from measured distributions as follows
 \begin{eqnarray}
 \label{eq:Dptsub}
 \Dptsub &=& \Dptmeas - \frac{\fd \cnchUE}{\fd \pTch}, \\
 \Dzsub  &=& \Dzmeas - \frac{ \fd \cnchUE}{\fd z}.
 \end{eqnarray}
While the correction for the UE can be large -- in the most 
central collisions the UE exceeds the signal by more than a 
factor of ten -- the correction for the presence of fake tracks is 
small, typically below 2\%.
   
The UE and fake-track-subtracted measured distributions, \Dptsub\ and 
\Dzsub , need to be corrected for resolution effects. There are two 
main resolution effects: smearing due to finite jet energy resolution and smearing due to 
finite track momentum resolution.
The former involves unfolding in \pTjet ; the 
latter involves unfolding in \pTch . Since the tracks were 
measured in jets, a two-dimensional unfolding needs to be used to 
correct for both of these resolution effects simultaneously. The 
two-dimensional Bayesian unfolding algorithm~\cite{DAgostini:1994zf} from the 
RooUnfold package~\cite{roounfold} was used for this purpose. Using the 
MC samples, four-dimensional response matrices were created using the 
truth and reconstructed \pTjet\ and the truth and reconstructed \pTch\ for 
reconstructed charged particles satisfying the track selection criteria 
defined in Section~\ref{sec:reconstruction}. The response matrices were created 
separately for \pp\ and \PbPb\ data for each centrality and 
rapidity range. The entries in the response matrix were weighted by 
the tracking efficiency correction. 
Five iterations in the Bayesian unfolding procedure were found 
sufficient to deliver a stable result that does not change with 
increasing numbers of iterations for all centrality bins except for 
the 0--10\% centrality bin where, eight iterations were used.
Once the two-dimensional distributions were unfolded, a projection 
to a given \pTjet\ interval was made, and the distribution was 
normalised by the respective number of jets.

The fragmentation distributions were measured for all jets 
reconstructed in the calorimeter, including those jets that do not 
contain any charged particle with $\pTch > 1~\GeV$.
The proper normalisation of the measured distributions by the number of jets 
requires a separate unfolding of the jet \pt\ spectrum. This was performed 
by applying a one-dimensional Bayesian unfolding, separately in each 
centrality and rapidity interval. One or two iterations were 
found to be sufficient for unfolding jet spectra in various centrality 
and rapidity intervals. The unfolded jet \pt\ spectra were 
integrated over a given jet \pt\ interval. The result of this 
integration represents the total number of jets spanning a given \pt\ 
interval and was used to normalise the unfolded fragmentation distributions,
\Dptunf\ and \Dzunf, as follows
 \begin{eqnarray}
 \label{eq:Dpt}
   D(\pt) &=& \frac{1}{N_\mathrm{jet}} \Dptunf, \\
   D(z) &=& \frac{1}{N_\mathrm{jet}} \Dzunf,
 \end{eqnarray}
where $D(\pt)$ and $D(z)$ are the final, particle-level corrected 
distributions that are presented in Section~\ref{sec:results}.

The performance of the reconstruction procedure was 
tested in MC samples by comparing unfolded distributions  
to truth distributions. 
Statistically independent MC samples for the response and reconstructed 
distributions were used.
The ratio of unfolded to truth 
distributions was found to be consistent with unity for all the bins 
used in the measurement.
An independent check of the subtraction of the UE contribution from measured
distributions was performed by estimating the UE charged-particle \pt\
spectra from the minimum-bias data sample. After applying centrality
reweighting, these UE charged-particle \pt\ spectra were found to be
consistent within statistical uncertainties with UE distributions
obtained by the cone method.
The performance of the unfolding procedure was further tested in 
the data by a procedure in which unfolded distributions were folded back 
using the response matrix. These ``refolded'' distributions were then 
compared to original raw distributions. Only differences at sub-percent level 
between the raw distributions and the refolded distributions 
were found.

%%%%%%%%%%%%%%%%%%%%%%%%%%%

\section{Systematic uncertainties}
\label{sec:systematics}

The following sources
of systematic uncertainty were identified for this measurement:
the uncertainties in the jet energy scale (JES) and 
jet energy resolution (JER), the track reconstruction efficiency,
and the unfolding.
The systematic uncertainties were evaluated separately for
distributions and their ratios for each
rapidity and centrality selection.

The systematic uncertainty due to the JES has two contributions: the \pp\ 
JES uncertainty and the heavy-ion JES uncertainty.
The impact of the JES uncertainty on the measured distributions was determined 
by shifting the transverse momentum of reconstructed jets as follows,
  \begin{equation}
  \pt' = \pt \cdot (1 \pm U^\mathrm{JES}(\pt,y)),
  \end{equation}
where $U^\mathrm{JES}(\pt,y)$ is either the \pp\ JES uncertainty \cite{Aad:2011he}
or centrality-dependent heavy-ion JES uncertainty \cite{HIjesnote}. The distributions with 
shifted \pt\ were unfolded and compared to the original distributions. 
The fractional difference was used as an estimate of the systematic 
uncertainty. The size of the JES uncertainty for $D(\pt)$ and $D(z)$ 
distributions in \pp\ collisions is typically below 2\% but can reach  
4\% and 6\% at high \pt\ and $z$, respectively. In Pb+Pb collisions, 
the typical size of this uncertainty is the same as in \pp\ collisions, 
but the maximal uncertainty can reach 15\% at the largest \pt\ or 
$z$. The JES uncertainty partially cancels in ratios of \PbPb\ and \pp\ 
distributions where a typical JES uncertainty is below 1\% and 
the maximal uncertainty is below 10\% at high \pt .
To account for systematic uncertainties due to any disagreement between 
the JER in data and MC simulation, the unfolding procedure was 
repeated with a modified response matrix. The new matrix was generated by 
repeating the MC study with the \pt\ of reconstructed jets smeared by a 
relative uncertainty estimated as a function of $y$ and \pt\ of the 
jet \cite{Aad:2011he}.
The size of the JER uncertainty is usually at the level of 1\% but grows at 
high \pt\ or $z$, where the maximum is $\approx 6\%$.

The systematic uncertainty due to track reconstruction 
was estimated by performing the analysis with three different sets of 
selection criteria imposed on tracks, called ``loose'', ``standard'', 
and ``tight''. The standard selection criteria were used as a default in 
this analysis. The differences in the result obtained using loose and 
tight criteria with respect to the result obtained using the standard 
criteria were used as the estimate of the systematic uncertainty.
The tight selection criteria imposed more stringent requirements on 
the track quality, leading to a 15--20\% reduction of the tracking 
efficiency depending on the track \pt , $\eta$, and centrality. The 
loose selection criteria imposed more relaxed requirements on track 
quality leading to a 5--10\% enhancement of tracking efficiency. The 
differences in the selection criteria bring significant differences both 
in the magnitude and the \pt dependence of the tracking efficiency. The 
track reconstruction uncertainty is usually largest systematic uncertainty at low and 
intermediate \pt\ or $z$. This uncertainty is typically less than
4\%. Also related to tracking are the uncertainty in the estimate of fake 
tracks and the uncertainty due to the parameterisation of tracking 
efficiencies. Both of these uncertainties are less than 2\%. 

The unfolding procedure is sensitive to the MC model and the number of 
iterations used, $N_\mathrm{it}$. Two variations were 
implemented to account for this systematic uncertainty. First, the $N_\mathrm{it}$ 
was varied by $\pm 1$. 
Second, the MC response matrix was reweighted such that its projection onto the reconstructed axis matches the data.
The data were then unfolded using the modified response matrix. 
The differences with respect to the original unfolded data were taken 
as the systematic uncertainty. The uncertainty due to unfolding was usually negligible and typically does not exceed 1\%. 
To determine the total systematic uncertainty, the systematic 
uncertainties from all different sources were added in quadrature.

%%%%%%%%%%%%%%%%%%%%%%%%%%%
\section{Results}
\label{sec:results}

The measurements of the internal structure of jets were performed 
differentially in jet \pt\ and $y$ and for  
two collision systems, \pp\ and \PbPb . In the case of \PbPb\ 
collisions, the measurement was performed in seven bins of 
centrality, 0--10\%, 10--20\%, 20--30\%, 30--40\%, 
40--50\%, 50--60\%, and 60-80\%.

\begin{figure*}
\centerline{
\includegraphics[width=0.89\textwidth]{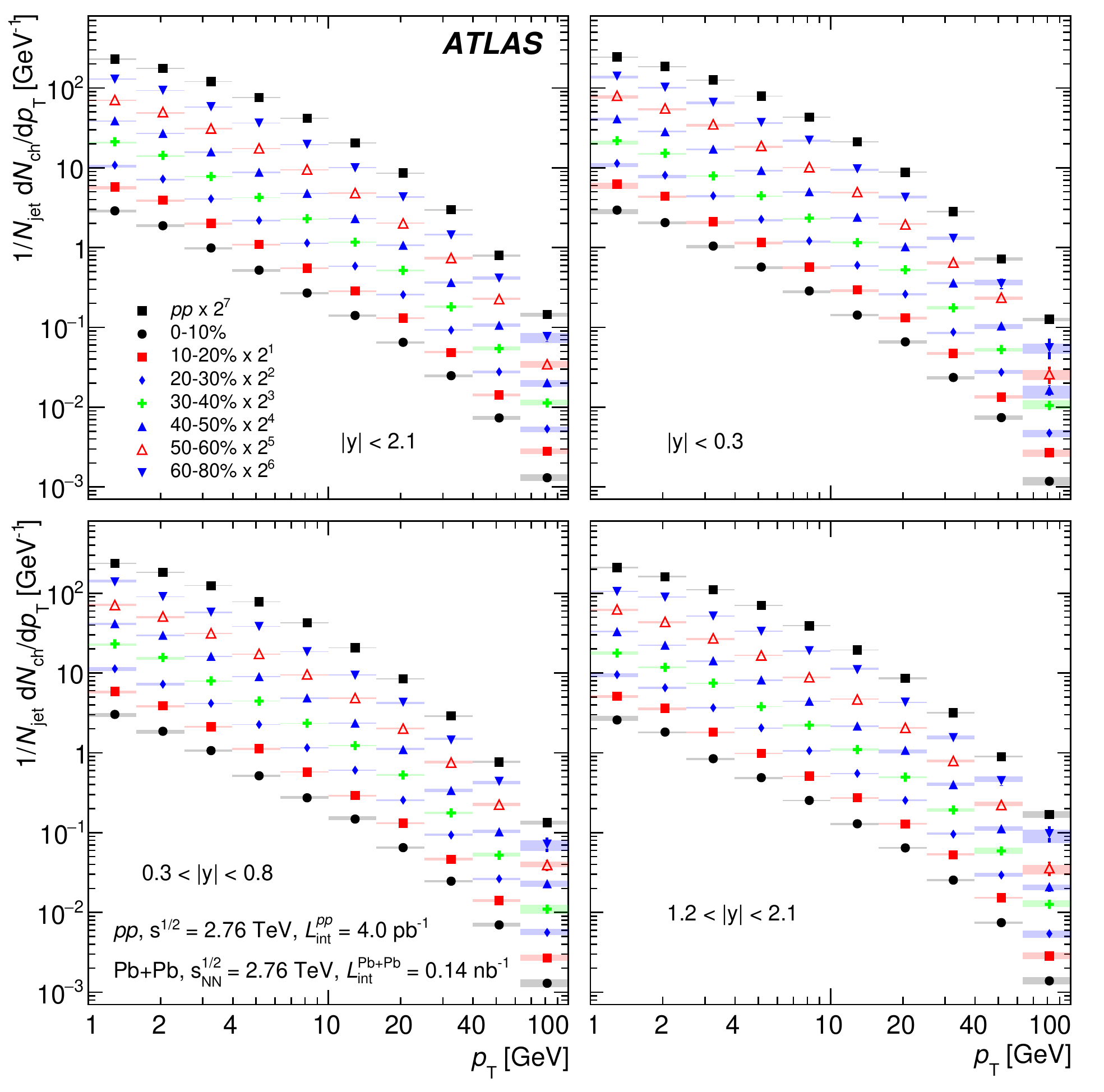}
}
\caption{
Unfolded charged-particle transverse momentum distributions, $D(\pt)$, 
measured in \pp\ collisions and for seven centrality bins measured in 
\PbPb\ collisions. The four panels show $D(\pt)$ distributions with 
different selections in jet rapidity for jets with \pt\ in the 
interval of 100--398~\GeV. 
The error bars on the data points indicate statistical uncertainties 
while the shaded bands indicate systematic uncertainties.
}
\label{fig:dist_dpt}
\end{figure*}
\begin{figure*}
\centerline{
\includegraphics[width=0.89\textwidth]{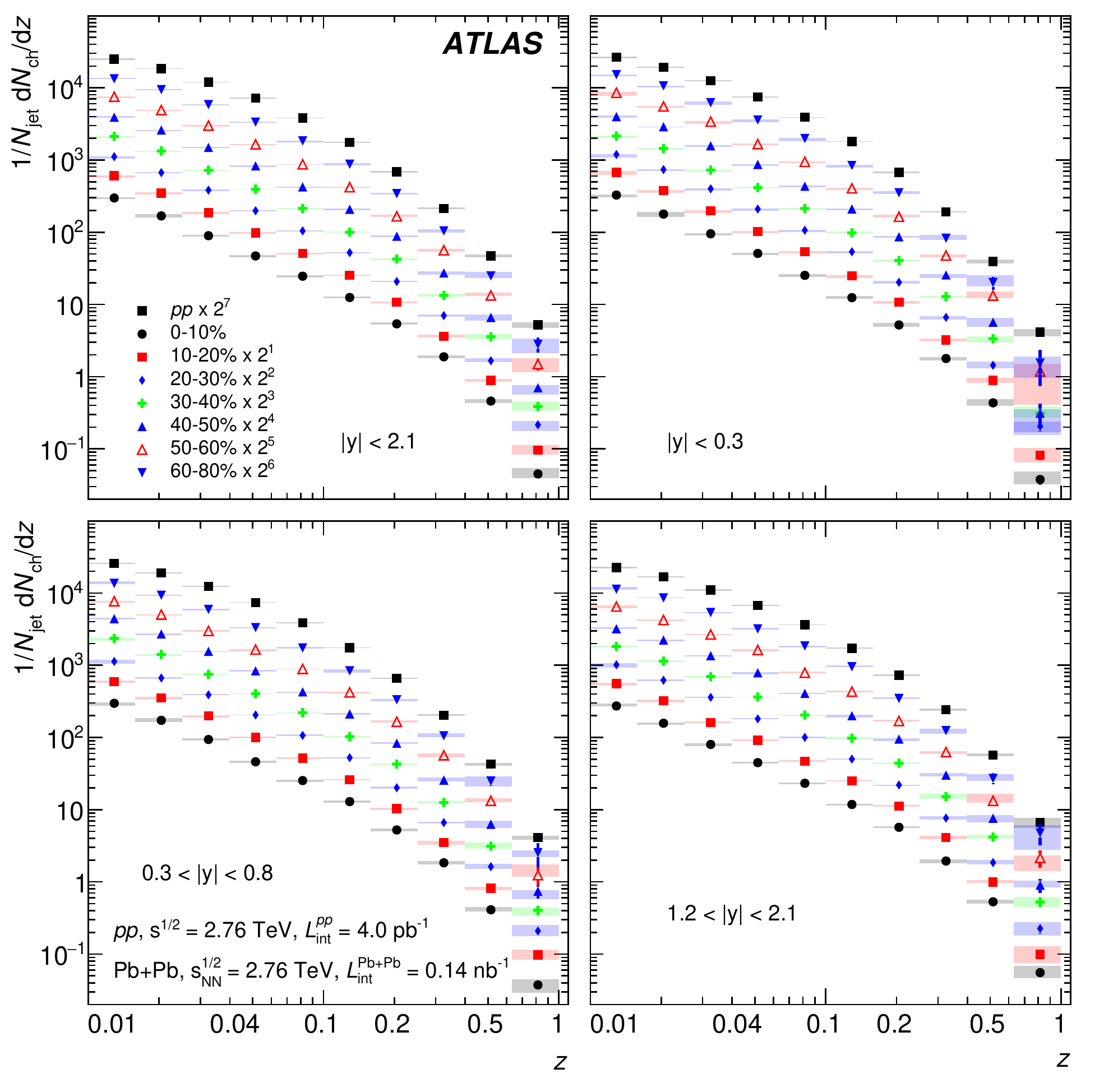}
}
\caption{
Unfolded distributions of longitudinal momentum fraction, $D(z)$, 
measured in \pp\ collisions and for seven centrality bins measured in 
\PbPb\ collisions. The four panels show $D(z)$ distributions with 
different selections in jet rapidity for jets with \pt\ in the 
interval of 100--398~\GeV. 
The error bars on the data points indicate statistical uncertainties 
while the shaded bands indicate systematic uncertainties.
}
\label{fig:dist_dz}
\end{figure*}

The measured distributions were evaluated in four different rapidity 
intervals of the jet: $|y|<2.1$, $|y|<0.3$, $0.3<|y|<0.8$, and 
$1.2<|y|<2.1$. 
The rapidity interval of $0.8<|y|<1.2$ was not
considered in the analysis
since the jet shape measurements are degraded in this region due to the
transition in the detector between the SCT barrel and end-caps.
This rapidity interval was also excluded from the measurement in
the full rapidity range, $|y|<2.1$.
The distributions were also evaluated in four different jet \pt\ 
intervals: $100 < \pTjet < 398~\GeV$, $100 < \pTjet < 126~\GeV$, $126 < 
\pTjet < 158~\GeV$, and $158 < \pTjet < 398~\GeV$. These intervals were 
chosen to correspond to intervals selected in the measurement of the jet 
nuclear modification factor \cite{Aad:2014bxa}. This should allow 
the size of the energy lost by a jet, as quantified by the nuclear modification factor,
to be connected to the respective modification of the jet fragmentation.

The $D(\pt)$ and $D(z)$ distributions corrected to the hadron level by the 
unfolding procedure described in Section~\ref{sec:analysis} are shown 
in Figures~\ref{fig:dist_dpt}~and~\ref{fig:dist_dz}, respectively. Different panels show 
distributions evaluated for different rapidity intervals for jets 
with $100 < \pt < 398~\GeV$. The shaded band represents the total systematic 
uncertainty, while the error bars represent statistical uncertainties. The 
distributions exhibit a difference in shape between central 
heavy-ion collisions and peripheral heavy-ion collisions or the \pp\ 
reference. To quantify this difference, the ratios of $D(\pt)$ and $D(z)$ 
distributions measured in heavy-ion collisions to those measured in \pp\ 
collisions were calculated and termed $R_{D(\pt)}$ and $R_{D(z)}$, 
respectively, following the nomenclature introduced in Ref.~\cite{Aad:2014wha},
  \begin{equation}
  R_{D(\pt)} = D(\pt)|_\mathrm{cent} / D(\pt)|_{pp}, ~~ 
  R_{D(z)} = D(z)|_\mathrm{cent} / D(z)|_{pp},
  \end{equation}
  where `cent' represents one of the seven centrality bins.

\begin{figure*}
\centerline{
\includegraphics[width=0.89\textwidth]{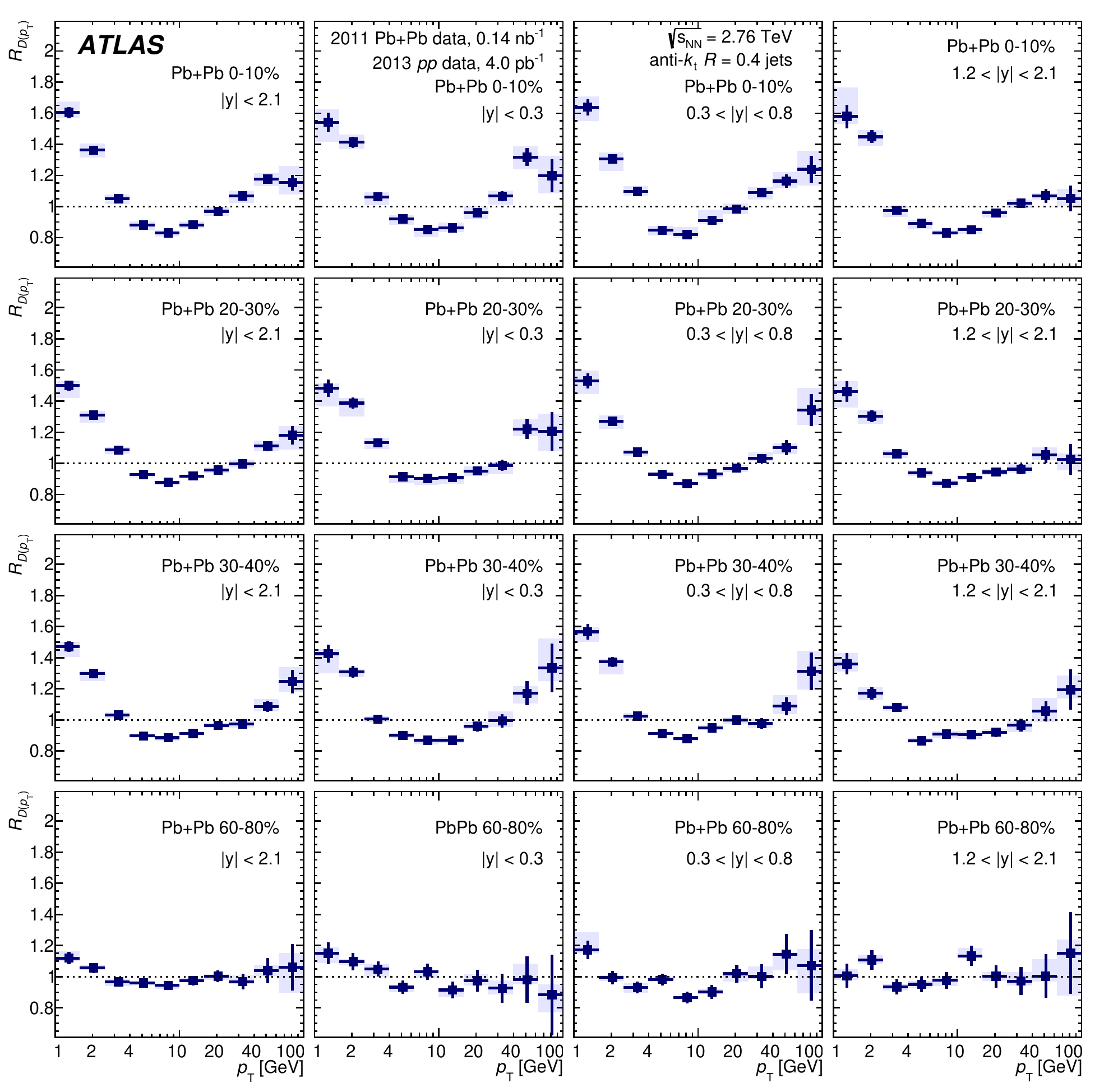}
}
\caption{
The ratio $R_{D(\pt)}$ of unfolded $D(\pt)$ distributions measured in heavy-ion 
collisions to unfolded $D(\pt)$ distributions measured in \pp\ collisions.
The $R_{D(\pt)}$ distributions were evaluated in four different centrality 
bins (rows) and four different selections in jet rapidity of jets (columns) with 
$100 < \pt < 398~\GeV$.
The error bars on the data points indicate statistical uncertainties 
while the shaded bands indicate systematic uncertainties.
}
\label{fig:rdpt_eta}
\end{figure*}

\begin{figure*}
\centerline{
\includegraphics[width=0.89\textwidth]{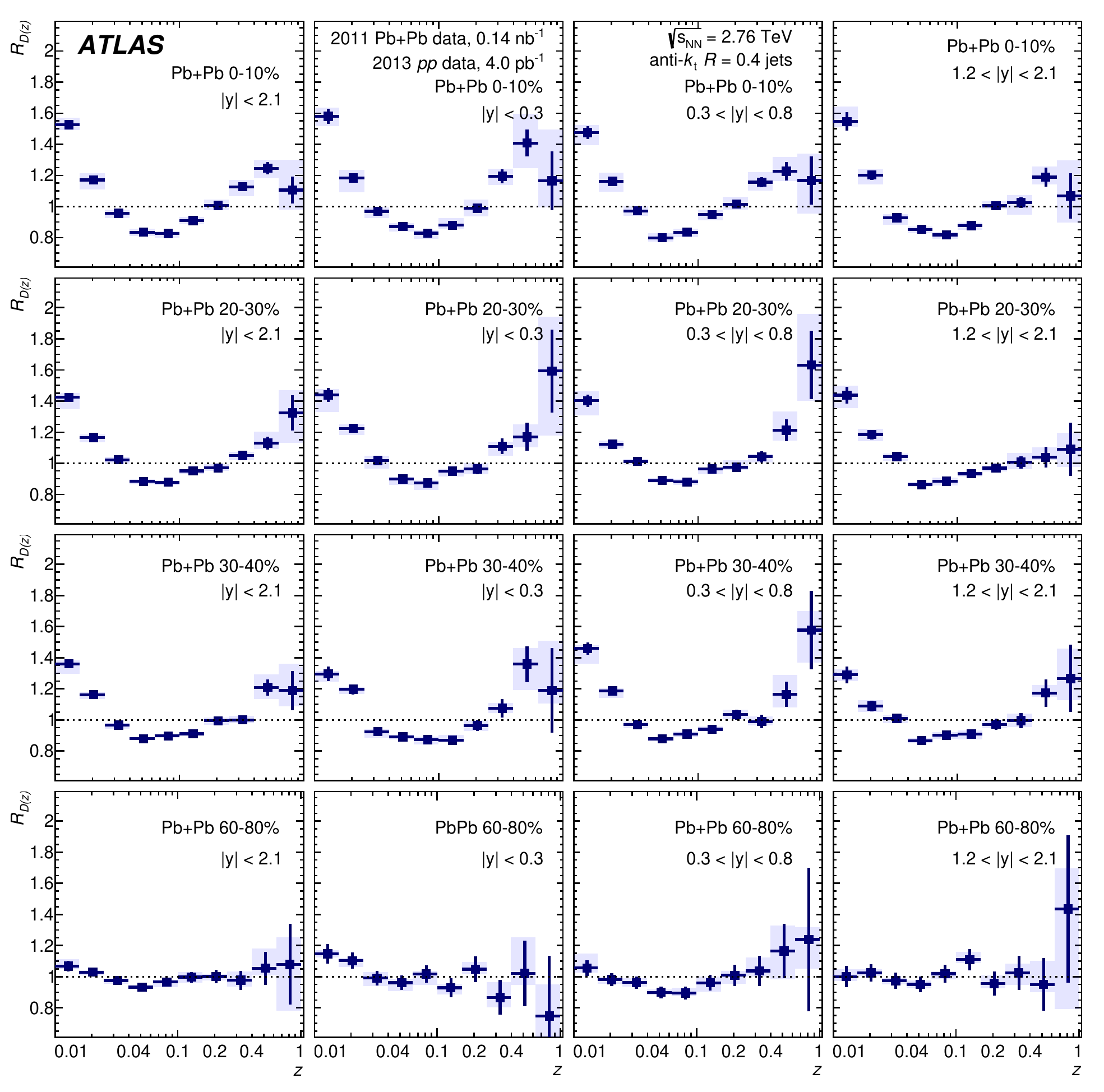}
}
\caption{
The ratio $R_{D(z)}$ of unfolded $D(z)$ distributions measured in heavy-ion 
collisions to unfolded $D(z)$ distributions measured in \pp\ collisions.
The $R_{D(z)}$ distributions were evaluated in four different centrality 
bins (rows) and four different selections in jet rapidity of jets (columns) with 
$100 < \pt < 398~\GeV$.
The error bars on the data points indicate statistical uncertainties 
while the shaded bands indicate systematic uncertainties.
}
\label{fig:rdz_eta}
\end{figure*}

\begin{figure*}
\centerline{
\includegraphics[width=0.89\textwidth]{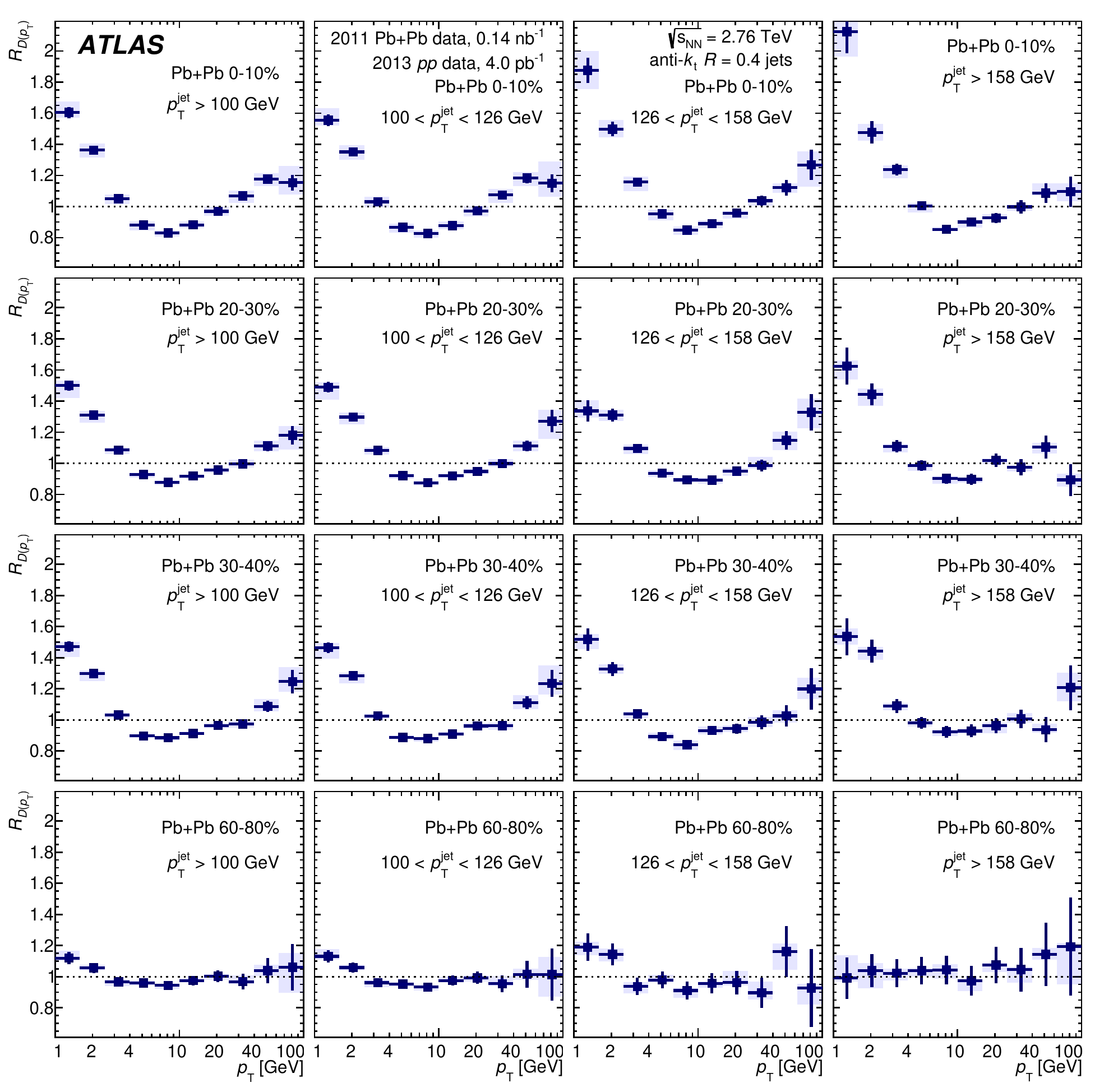}
}
\caption{
The ratio $R_{D(\pt)}$ of unfolded $D(\pt)$ distributions measured in heavy-ion 
collisions to unfolded $D(\pt)$ distributions measured in \pp\ collisions.
The $R_{D(\pt)}$ distributions were evaluated in four different centrality 
bins (rows) and four different selections in jet \pt\ of jets (columns) with 
$|y|<2.1$.
The error bars on the data points indicate statistical uncertainties 
while the shaded bands indicate systematic uncertainties.
}
\label{fig:rdpt_pt}
\end{figure*}

\begin{figure*}
\centerline{
\includegraphics[width=0.89\textwidth]{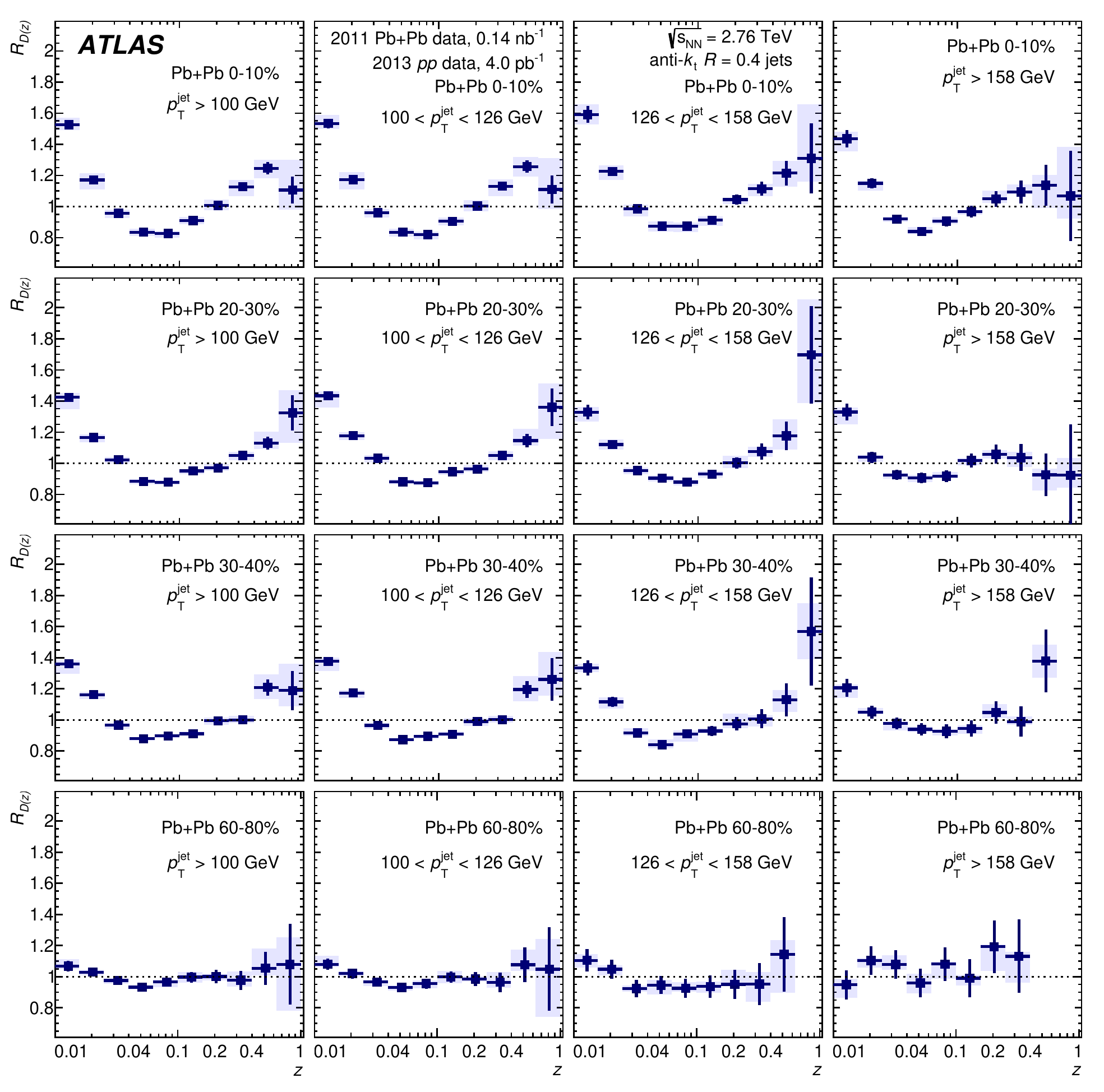}
}
\caption{
The ratio $R_{D(z)}$ of unfolded $D(z)$ distributions measured in heavy-ion 
collisions to unfolded $D(z)$ distributions measured in \pp\ collisions.
The $R_{D(z)}$ distributions were evaluated in four different centrality 
bins (rows) and four different selections in jet \pt\ of jets (columns) with 
$|y|<2.1$.
The error bars on the data points indicate statistical uncertainties 
while the shaded bands indicate systematic uncertainties.
}
\label{fig:rdz_pt}
\end{figure*}

The $R_{D(\pt)}$ and $R_{D(z)}$ distributions are shown in 
Figures~\ref{fig:rdpt_eta}--\ref{fig:rdz_pt}. Figure~\ref{fig:rdpt_eta} 
shows the $R_{D(\pt)}$ distributions for four selections in collision 
centrality, namely 0--10\%, 20--30\%, 30--40\% and 60--80\%, and for 
four rapidity intervals of jets with \pTjet\ in the interval of 
100--398~\GeV.
These ratios show an enhancement in fragment yield in central 
collisions for $\pTch<4~\GeV$, a reduction in fragment yields for 
$4<\pTch<25~\GeV$, and an enhancement in the fragment yield for $\pTch > 
25~\GeV$. 
The magnitude of these modifications decreases for more peripheral 
collisions.
A similar observation is 
also made for the $R_{D(z)}$ distributions shown in Figure~\ref{fig:rdz_eta}.
The characteristic shape of these ratios was also seen in the 
previous study \cite{Aad:2014wha} where the 60--80\% bin was used as a reference.
Figures~\ref{fig:rdpt_eta} and ~\ref{fig:rdz_eta} show that the 
difference in the modifications between different rapidity 
selections is marginal for fragments with $\pTch<25~\GeV$ and $z<0.25$, 
respectively. Only at high \pTch\ or high $z$, 
a small difference is observed -- the enhancement is 
systematically lower for more forward jets than for jets measured in 
the central rapidity region.

Figures~\ref{fig:rdpt_pt} and \ref{fig:rdz_pt} show the $R_{D(\pt)}$ and 
$R_{D(z)}$ distributions, respectively, both for four \pTjet\ intervals 
of jets with $|y|<2.1$. No significant differences can be observed among 
the four \pTjet\ selections.

%%%%%%%%%%%%%%%%%%%%%%%%%%%
\section{Discussion}
\label{sec:discussion}

  To quantify the trends seen in the ratios, the differences between 
integrals of $D(\pt)$ distributions measured in heavy-ion collisions and
the integrals of $D(\pt)$ distributions measured in \pp\ collisions, \NUM , were 
evaluated,
  \begin{equation}
  \NUM|_\mathrm{cent} \equiv \int_{\ptmin}^{\ptmax} \bigg( D(\pt)|_\mathrm{cent} - D(\pt)|_{\pp} \bigg) ~ \fd \pt .
  \end{equation}
Three ranges 
defined by values of \ptmin\ and \ptmax\ were chosen 
to match the observations in $R_{D(\pt)}$, namely 
1--4~\GeV, 4--25~\GeV, and 25--100~\GeV.
Thus three values of \NUM\ were obtained for each centrality bin which 
represent the number of particles carrying: (1) the excess seen in heavy-ion 
collisions for particles with $1 < \pt < 4~\GeV$, (2) a depletion 
seen for particles with $4 < \pt < 25~\GeV$, and (3) the enhancement seen 
for particles with $25 < \pt < 100~\GeV$.
Further, the differences between integrals of the first moment of the 
$D(\pt)$ distributions, \MOM , were also evaluated,
  \begin{equation}
  \label{eqn:MOM}
  \MOM|_\mathrm{cent} \equiv \int_{\ptmin}^{\ptmax} \bigg( D(\pt)|_\mathrm{cent} - D(\pt)|_{\pp} \bigg) ~ \pt ~ \fd \pt .
  \end{equation}
These differences represent the total transverse momentum of 
particles carrying the excess or the depletion observed in $R_{D(\pt)}$ 
distributions.

\begin{figure*}
\centerline{
\includegraphics[width=0.98\textwidth]{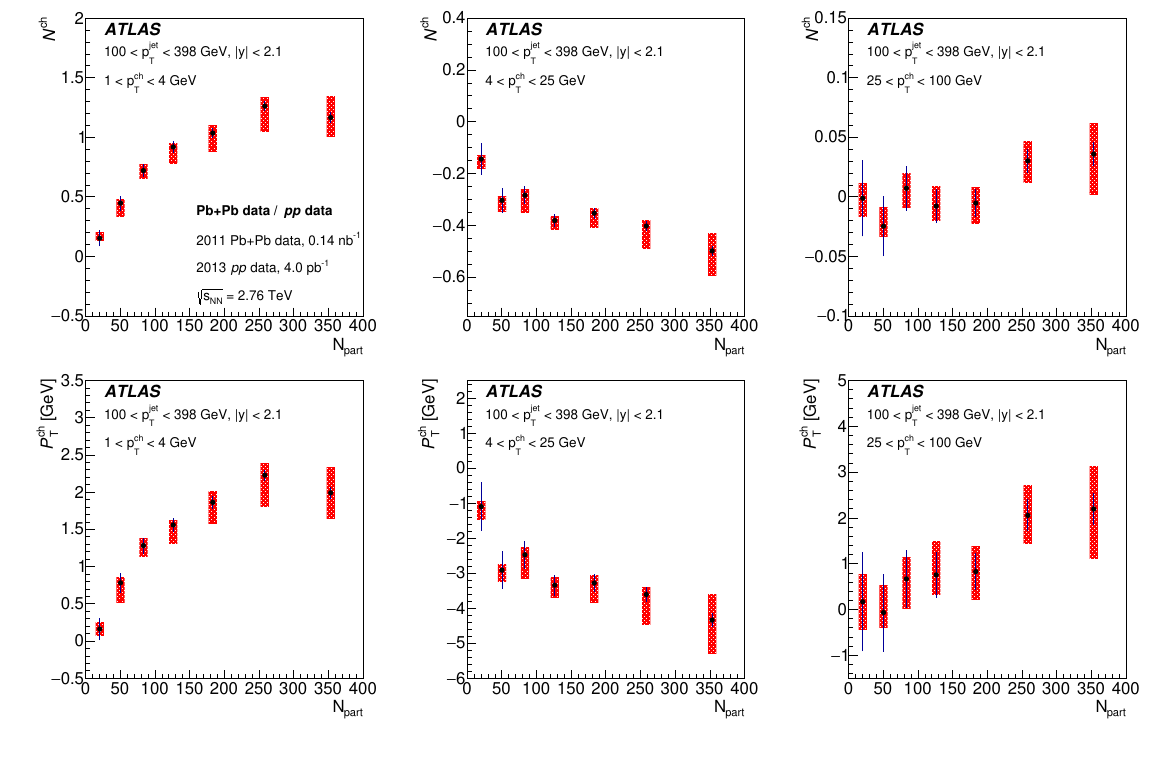}
}
\caption{
(Upper panels) The difference \NUM\ between the total yield of particles in a given 
\pTch\ interval (indicated in the legend) measured in \PbPb\ collisions and the 
total yield of particles in the same \pTch\ interval measured in \pp\ collisions. 
(Lower panels) The difference \MOM\ between the total transverse momentum of 
particles in a given \pTch\ interval measured in \PbPb\ collisions and the total 
transverse momentum of particles measured in \pp\ collisions. The differences were 
evaluated as a function of number of participating nucleons, $\Npart$.
The error bars on the data points indicate statistical uncertainties 
while the shaded bands indicate systematic uncertainties.
}
\label{fig:res_integrals}
\end{figure*}

The result of performing this calculation is shown in 
Figure~\ref{fig:res_integrals} where the differences between the two integrals are 
displayed as a function of the number of participants, $\Npart$, calculated using the 
Glauber model analysis of the \ETfcal\ 
\cite{ATLAS:2011ah,Alver:2008aq,Miller:2007ri}.
A clear, almost logarithmic, increase of yields of particles with low 
transverse momenta with increasing centrality is seen. In contrast, the 
intermediate-\pTch\ region exhibits less significant modifications 
with varying centrality. The yield at high \pTch\ 
shows a mild increase with increasing centrality, however with smaller significance. 
The changes in the total transverse momentum follow the trends seen in 
\Rdpt\ distributions. However, for the high-\pt\ region, the 
significance of the increase in yields is more pronounced in \Rdpt\ 
distributions than in the $\MOM$ distribution.

The difference defined in Eq.~(\ref{eqn:MOM}) can also be evaluated over the 
full range of charged-particle transverse momenta, $1 < \pTch < 
100~\GeV$. It may be expected that such \MOM\ should be identical to zero 
since the same range of the \pTjet\ was used in \PbPb\ and \pp\ 
collisions. The result of this evaluation is presented in the second row 
of Table~\ref{tab:tab1}. Indeed, the \MOM\ evaluated over the full range 
of charged-particle transverse momenta is consistent with zero within 
one standard deviation of combined statistical and systematic 
uncertainties. The small residual deviations from zero are likely due 
to the difference in the shape of \pTjet\ spectra between \pp\ and 
\PbPb\ collisions \cite{Aad:2014bxa}, which leads to a difference in the 
mean \pTjet\ between \PbPb\ and \pp\ collisions.

The total difference in the yield of charged particles can 
also be evaluated by integrating the $D(\pt)$ distributions over the full 
range of charged-particle transverse momenta. In this case, one does not 
expect to see the same yields of charged particles in \PbPb\ and \pp\ 
collisions since this quantity may change as a result of the jet 
quenching. The resulting \NUM\ is summarised in the bottom row of 
Table~\ref{tab:tab1}.

\begin{table*}
\begin{center}
\begin{tabular}{|l|r@{}l|r@{}l|r@{}l|r@{}l|r@{}l|r@{}l|r@{}l|}
\hline
Centrality            & \multicolumn{2}{c|}{0--10\%} & \multicolumn{2}{c|}{10--20\%} & \multicolumn{2}{c|}{20--30\%} & \multicolumn{2}{c|}{30--40\%} & \multicolumn{2}{c|}{40--50\%} & \multicolumn{2}{c|}{50--60\%} & \multicolumn{2}{c|}{60--80\% }\\ \hline

\MOM\ [\GeV]           & $  0.9$&$^{ + 0.9}_{ - 1.7}$
                      & $  1.0$&$^{ + 0.8 }_{ - 1.3}$
                      & $ -0.0$&$^{ + 0.7}_{ - 1.1}$
                      & $ -0.6$&$^{ + 0.8}_{ - 0.8} $
                      & $ -0.5$&$^{ + 1.0}_{ - 1.2} $
                      & $ -1.4$&$^{ + 1.0}_{ - 1.2} $
                      & $ -0.8$&$^{ + 1.3}_{ - 1.4} $ \\
\hline
\NUM\                 & $ 0.7$&$^{ + 0.1}_{ -0.2 }$
                      & $ 0.9$&$^{ + 0.1}_{ -0.1 } $
                      & $ 0.7$&$^{ + 0.1}_{ -0.1 } $
                      & $ 0.5$&$^{ + 0.1}_{ -0.2 }$
                      & $ 0.4$&$^{ + 0.1}_{ -0.1 }$
                      & $ 0.2$&$^{ + 0.1}_{ -0.2 } $
                      & $ 0.0$&$^{ + 0.1}_{ -0.1 }$ \\
\hline
\end{tabular}
\end{center}
\caption{
  The difference between \pp\ and \PbPb\ collisions in the total momentum, \MOM , and the total difference in
the yield of charged particles between \pp\ and \PbPb\ collisions, \NUM , evaluated over the full range of
charged-particle transverse momenta, $1 < \pTch < 100~\GeV$, and for different values of centrality.
  }
\label{tab:tab1}
\end{table*}

\begin{figure*}
\centerline{
\includegraphics[width=0.98\textwidth]{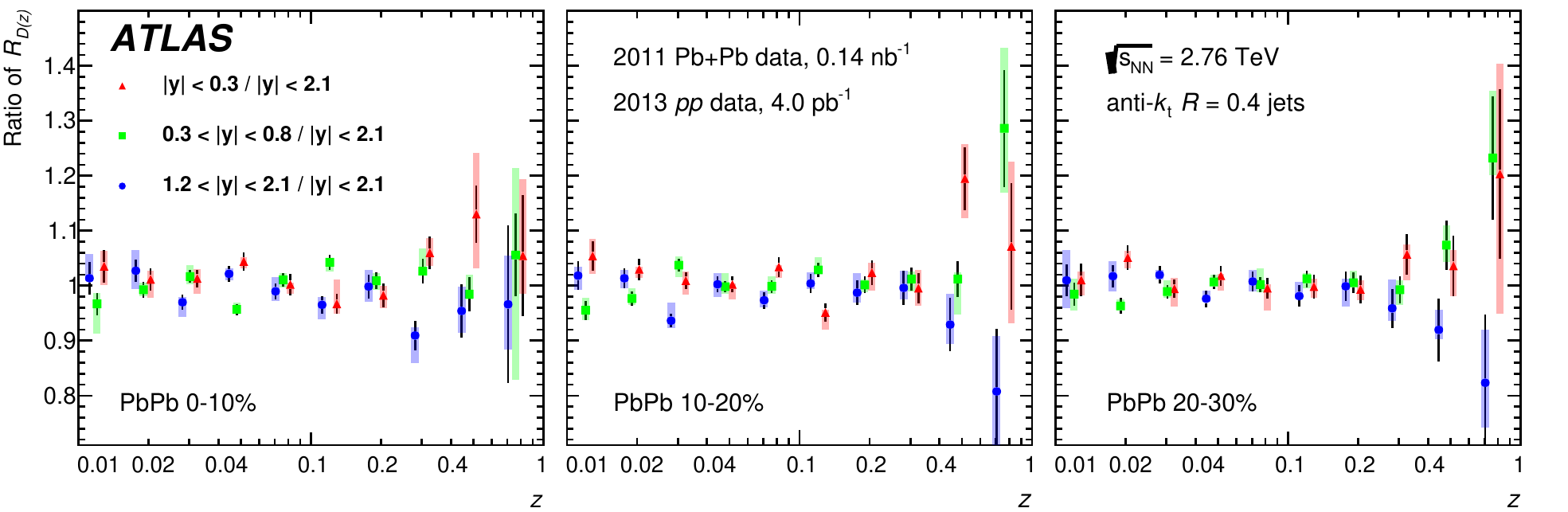}
}
\caption{
  The ratio of \Rdz\ distributions in a given rapidity interval, namely 
$|y|<0.3$, $0.3<|y|<0.8$, and $1.2<|y|<2.1$, to \Rdz\ in $|y|<2.1$. The 
ratio of \Rdz\ was evaluated for three different collision centralities. 
The error bars on the data points indicate statistical uncertainties 
while the shaded bands indicate systematic uncertainties.
  }
\label{fig:double}
\end{figure*}

The enhancement of fragment yields at low \pt\ or $z$ already reported in 
previous analyses \cite{Aad:2014wha,Chatrchyan:2014ava} is confirmed, and it 
is consistent with a jet quenching interpretation
in which the energy lost by 
partons is transferred predominantly to soft particles 
\cite{Chatrchyan:2011sx}. 
While the enhancement of soft fragments may be understood as a direct 
consequence of the parton energy loss,
the enhancement of fragment yields at high \pt\ or $z$ is unexpected 
\cite{Abreu:2007kv}. A discussion of this feature 
in terms of different quenching of quark and gluon jets
was recently provided 
in Ref.~\cite{Spousta:2015fca}. 
In order to further study this enhancement the ratio of \Rdz\ 
distributions in a given rapidity interval to \Rdz\ in $|y|<2.1$ is 
evaluated and plotted in Figure~\ref{fig:double}. At high $z$ ($z \gtrsim 
0.4$) the result shows a trend of enhancements in the ratio of \Rdz\ 
measured in $|y|<0.3$ to \Rdz\ in $|y|<2.1$ and a trend of depletions
in the ratio of \Rdz\ measured in $1.2<|y|<2.1$ to \Rdz\ in $|y|<2.1$. 

%%%%%%%%%%%%%%%%%%%%%%%%%%%
\clearpage
\newpage
\section{Summary}
\label{sec:summary}

This paper presents a measurement of internal structure of jets
performed with the ATLAS detector at the LHC. The distributions of
charged-particle transverse momentum and longitudinal momentum fraction are
measured in jets reconstructed using the \antikt\ algorithm with
$R=0.4$. These distributions are measured differentially in jet \pt ,
jet rapidity, and in \PbPb\ as well as \pp\ collisions at a centre-of-mass energy of 2.76 TeV per colliding nucleon pair. 
The Pb+Pb and \pp\ data correspond to integrated luminosities of 0.14~nb$^{-1}$ and 4.0~pb$^{-1}$, respectively.
In the case of \PbPb\ collisions, the measurements are performed in bins of 
collision centrality.
The distributions
measured in \pp\ collisions are used as a reference for the
distributions measured in \PbPb\ collisions to evaluate the impact of the
jet energy loss on the internal structure of jets.
The measurements cover the jet \pt\ range of 100--398~\GeV\ and use charged 
particles with $\pt >1~\GeV$. The results are corrected to the hadron level.

The ratios of charged-particle transverse momentum distributions
measured in \PbPb\ collisions to those measured in \pp\ exhibit an
enhancement in fragment yield in central collisions for $1<\pTch<4~\GeV$, a
reduction in fragment yields for $4<\pTch<25~\GeV$, and an enhancement in
the fragment yield for $\pTch > 25~\GeV$. 
The magnitude of these modifications
decreases in more peripheral collisions. 
A similar observation is also made for the distributions of 
longitudinal momentum fraction measured with respect to the jet axis.

The centrality dependence of the magnitude of modifications was further
quantified by evaluating the differences between integrals of charged-particle 
transverse momentum distributions measured in \PbPb\ and \pp\
collisions for these three characteristic \pTch\ intervals.
Further, the jet \pt - and $y$-dependence of the modifications in
the internal structure of jets was measured. 
In addition, no significant differences in modifications of the jet
structure are observed among different \pTjet\ selections spanning 
the interval of 100--398~\GeV.
No significant evolution in modifications of the jet
structure as a function of rapidity are observed except for a difference 
at high \pTch\ or high $z$, where a hint of reduction of the enhancement for 
more forward jets is observed. 

These new results improve our understanding of the in-medium
modifications of parton showers and help to constrain jet-quenching models.

%-------------------------------------------------------------------------------
\section*{Acknowledgements}
%-------------------------------------------------------------------------------

% Acknowledgements for papers with collision data
% Version 06-Mar-2017

% Standard acknowledgements start here
%----------------------------------------------
We thank CERN for the very successful operation of the LHC, as well as the
support staff from our institutions without whom ATLAS could not be
operated efficiently.

We acknowledge the support of ANPCyT, Argentina; YerPhI, Armenia; ARC, Australia; BMWFW and FWF, Austria; ANAS, Azerbaijan; SSTC, Belarus; CNPq and FAPESP, Brazil; NSERC, NRC and CFI, Canada; CERN; CONICYT, Chile; CAS, MOST and NSFC, China; COLCIENCIAS, Colombia; MSMT CR, MPO CR and VSC CR, Czech Republic; DNRF and DNSRC, Denmark; IN2P3-CNRS, CEA-DSM/IRFU, France; SRNSF, Georgia; BMBF, HGF, and MPG, Germany; GSRT, Greece; RGC, Hong Kong SAR, China; ISF, I-CORE and Benoziyo Center, Israel; INFN, Italy; MEXT and JSPS, Japan; CNRST, Morocco; NWO, Netherlands; RCN, Norway; MNiSW and NCN, Poland; FCT, Portugal; MNE/IFA, Romania; MES of Russia and NRC KI, Russian Federation; JINR; MESTD, Serbia; MSSR, Slovakia; ARRS and MIZ\v{S}, Slovenia; DST/NRF, South Africa; MINECO, Spain; SRC and Wallenberg Foundation, Sweden; SERI, SNSF and Cantons of Bern and Geneva, Switzerland; MOST, Taiwan; TAEK, Turkey; STFC, United Kingdom; DOE and NSF, United States of America. In addition, individual groups and members have received support from BCKDF, the Canada Council, CANARIE, CRC, Compute Canada, FQRNT, and the Ontario Innovation Trust, Canada; EPLANET, ERC, ERDF, FP7, Horizon 2020 and Marie Sk{\l}odowska-Curie Actions, European Union; Investissements d'Avenir Labex and Idex, ANR, R{\'e}gion Auvergne and Fondation Partager le Savoir, France; DFG and AvH Foundation, Germany; Herakleitos, Thales and Aristeia programmes co-financed by EU-ESF and the Greek NSRF; BSF, GIF and Minerva, Israel; BRF, Norway; CERCA Programme Generalitat de Catalunya, Generalitat Valenciana, Spain; the Royal Society and Leverhulme Trust, United Kingdom.

The crucial computing support from all WLCG partners is acknowledged gratefully, in particular from CERN, the ATLAS Tier-1 facilities at TRIUMF (Canada), NDGF (Denmark, Norway, Sweden), CC-IN2P3 (France), KIT/GridKA (Germany), INFN-CNAF (Italy), NL-T1 (Netherlands), PIC (Spain), ASGC (Taiwan), RAL (UK) and BNL (USA), the Tier-2 facilities worldwide and large non-WLCG resource providers. Major contributors of computing resources are listed in Ref.~\cite{ATL-GEN-PUB-2016-002}.
%----------------------------------------------

%-------------------------------------------------------------------------------
%\clearpage
%\appendix
%\part*{Appendix}
%\addcontentsline{toc}{part}{Appendix}
%-------------------------------------------------------------------------------

%In a paper, an appendix is used for technical details that would otherwise disturb the flow of the paper.
%Such an appendix should be printed before the Bibliography.

%-------------------------------------------------------------------------------
% If you use biblatex and either biber or bibtex to process the bibliography
% just say \printbibliography here
\printbibliography
% If you want to use the traditional BibTeX you need to use the syntax below.
% \bibliographystyle{bibtex/bst/atlasBibStyleWoTitle}
% \bibliography{,bibtex/bib/ATLAS}
%-------------------------------------------------------------------------------

\newpage 
% ATLAS Collaboration author list
% Data extracted on 02-Feb-2017 for paper reference HION-2015-05
%\documentclass[11pt]{article}
%\usepackage{a4wide}\begin{document}
\begin{flushleft}
{\Large The ATLAS Collaboration}

\bigskip

M.~Aaboud$^\textrm{\scriptsize 137d}$,
G.~Aad$^\textrm{\scriptsize 88}$,
B.~Abbott$^\textrm{\scriptsize 115}$,
J.~Abdallah$^\textrm{\scriptsize 8}$,
O.~Abdinov$^\textrm{\scriptsize 12}$$^{,*}$,
B.~Abeloos$^\textrm{\scriptsize 119}$,
S.H.~Abidi$^\textrm{\scriptsize 161}$,
O.S.~AbouZeid$^\textrm{\scriptsize 139}$,
N.L.~Abraham$^\textrm{\scriptsize 151}$,
H.~Abramowicz$^\textrm{\scriptsize 155}$
% ...
\bigskip
\\
% ...
$^{8}$ Department of Physics, The University of Texas at Arlington, Arlington TX, United States of America\\
% ...
$^{12}$ Institute of Physics, Azerbaijan Academy of Sciences, Baku, Azerbaijan\\
% ...
$^{88}$ CPPM, Aix-Marseille Universit{\'e} and CNRS/IN2P3, Marseille, France\\
% ...
$^{115}$ Homer L. Dodge Department of Physics and Astronomy, University of Oklahoma, Norman OK, United States of America\\
% ...
$^{119}$ LAL, Univ. Paris-Sud, CNRS/IN2P3, Universit{\'e} Paris-Saclay, Orsay, France\\
% ...
$^{137}$ $^{(a)}$ Facult{\'e} des Sciences Ain Chock, R{\'e}seau Universitaire de Physique des Hautes Energies - Universit{\'e} Hassan II, Casablanca; $^{(b)}$ Centre National de l'Energie des Sciences Techniques Nucleaires, Rabat; $^{(c)}$ Facult{\'e} des Sciences Semlalia, Universit{\'e} Cadi Ayyad, LPHEA-Marrakech; $^{(d)}$ Facult{\'e} des Sciences, Universit{\'e} Mohamed Premier and LPTPM, Oujda; $^{(e)}$ Facult{\'e} des sciences, Universit{\'e} Mohammed V, Rabat, Morocco\\
% ...
$^{139}$ Santa Cruz Institute for Particle Physics, University of California Santa Cruz, Santa Cruz CA, United States of America\\
% ...
$^{151}$ Department of Physics and Astronomy, University of Sussex, Brighton, United Kingdom\\
% ...
$^{155}$ Raymond and Beverly Sackler School of Physics and Astronomy, Tel Aviv University, Tel Aviv, Israel\\
% ...
$^{161}$ Department of Physics, University of Toronto, Toronto ON, Canada\\
% ...
$^{a}$ Also at Department of Physics, King's College London, London, United Kingdom\\
$^{b}$ Also at Institute of Physics, Azerbaijan Academy of Sciences, Baku, Azerbaijan\\
$^{c}$ Also at Novosibirsk State University, Novosibirsk, Russia\\
$^{d}$ Also at TRIUMF, Vancouver BC, Canada\\
$^{e}$ Also at Department of Physics {\&} Astronomy, University of Louisville, Louisville, KY, United States of America\\
% ...
$^{*}$ Deceased
\end{flushleft}

\end{document}